\documentclass[10pt,journal]
{IEEEtran}

\usepackage{amsmath,epsfig}
\usepackage{amsfonts,amssymb}
\usepackage{fixmath}

\usepackage{color}

\ifCLASSINFOpdf

\else

\fi

\begin{document}
\onecolumn

\title{Multiview Navigation based on Extended Layered Depth Image Representation}

\author{Uday Takyar, Thomas~Maugey,
        and~Pascal~Frossard}

\maketitle

\begin{abstract}
Emerging applications in multiview streaming look for providing interactive navigation services to video players. The user can ask for information from any viewpoint with a minimum transmission delay. The purpose is  to provide user with as much information as possible with least number of redundancies. The recent concept of navigation segment representation consists of regrouping a given number of viewpoints in one signal and transmitting them to the users according to their navigation path. The question of the best description strategy of these navigation segments is however still open. In this paper, we propose to represent and code navigation segments by a method that extends the recent layered depth image (LDI) format. It consists of describing the scene from a viewpoint with multiple images organized in layers corresponding to the different levels of occluded objects. The notion of extended LDI comes from the fact that the size of this image is adapted to take into account the sides of the scene also, in contrary to classical LDI. The obtained results show a significant rate-distortion gain compared to classical multiview compression approaches in navigation scenario.

\end{abstract}


\section{Introduction}
There has been recently a lot of interest in 3D video communications, in particular with the advent of 3D-TV and Free viewpoint TV. Interactivity, which is the ability to explore and navigate audiovisual scenes by freely choosing viewpoint and viewing direction, is an important key feature in new and emerging audiovisual applications \cite{Gelman_A_2011_spie_cen_icmi}. The most common case of interactive services is free viewpoint video. In this case, a 3D scene is captured by multiple cameras from different viewpoints. In addition to the video signals, other information such as camera calibration and scene geometry is also acquired or estimated. This input data enables interactive navigation through the 3D scene where the user can choose to see any camera view or any synthetic view that can be reconstructed from camera images and  geometry information. 

\begin{figure}[htb]
 \centering
\centerline{\epsfig{figure= 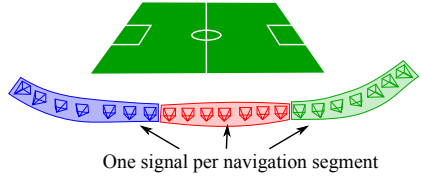,width=0.3\linewidth}}
\caption{The navigation domain is divided into navigation segments, which regroup multiple viewpoints described by one signal.}
\label{fig:NS}
\end{figure}

It becomes important to design an effective representation of the input data, which does not only provide good compression performance but also flexibility for navigation. There are various 3D representation methods available between image-based approaches and classical 3D computer graphics approaches, which rely on different forms of geometry information. Image-based methods work well if the data is very densely sampled but capturing such high quality data is quite a difficult task. In purely geometry-based representations, the reconstruction of real life scenes is quite complex and time consuming and the content creation is quite costly. Between these two families of representations, depth-based representations are trying to take the best of both worlds, by minimizing the need for dense image sampling, and avoiding the need for expensive content creation. For that, they provide gray-scale images depicting the distance between the scene points and the camera plane. Depth-based representations are typically very redundant and pose several challenges in designing effective and flexible compression schemes.
The classical multi-view video transmission schemes \cite{Wiegand_T_2003_tcsvt_ove_hvcs} transmit all the frames to the decoder and hence are not very effective for interactive systems with transmission constraints, since the decoder is generally interested by one view at a time only. In other words, classical multi-view video transmission schemes exploit correlation between views at the coding step, but leads to suboptimal results due to a priori unknown user navigation patterns. The work in \cite{Maugey_T_2013_tip_nav_dpimi} defines a concept of navigation domain and navigation segment in order to control the inter-view dependencies in the original data representation, for interactive applications. A navigation domain includes information about all possible viewpoints in a certain region and permits to synthesize any virtual views  in this navigation domain. The navigation domain is further divided into navigation segments, which represent independent entities for user navigation (see Fig.~\ref{fig:NS}). While navigation segments offer interesting perspectives in interactive services, they necessitate an effective data representation format, which contains information for virtual view reconstruction in a compact form. In this work, we propose a new representation strategy for efficiently describing the navigation segments.

In more details, we propose in this paper a data representation method that extends the notion of layered depth image (LDI) representations \cite{Cheng_CM_2008_icpr_imp_nvsfdilb,Karlsson_LS_2011_tcsvt_lay_abdddmpdsvc, Gelman_A_2012_tip_mul_icudloba}. A LDI represents a 3D scene with an array of pixels viewed from a single camera location. Each layered depth image pixel is represented by its color, depth and a few other properties necessary for rendering. There may be multiple pixels along each line of sight, hence the layered depth image usually consists of multiple levels of information. Then, only one layered depth image corresponding to several original views is constructed where all the common points between several views are warped to a single location. It enables the rendering of virtual views of the object/scene at new camera positions, and the rendering operation can be performed quickly with a warp-order algorithm \cite{Cheng_CM_2008_icpr_imp_nvsfdilb}. 
We build on the LDI format and propose a new representation that is adapted to navigation segments. The original LDI representation unfortunately discards the extra information corresponding to pixel positions that fall outside the boundaries of the original image after warping to the LDI camera view. We rather incorporate this information  by extending the size of layered depth image beyond the original image size, so that all the information corresponding to a navigation segment, i.e., each 3D point of the scene, is represented once and only once. Therefore, the proposed extended LDI removes any inter-view redundancy in the 3D scene description. 

Since we build a representation that represents the 3D scene with simple image juxtaposition, we then use an adaptation of the LDI compression technique \cite{Duan_J_2003_tip_com_ldi} for coding the navigation segment. It consists of image compression tools that are adapted to the LDI type structure. We finally show by experiments that our new format outperforms classical tools such as MVC coder \cite{jmvm} and provides an effective solution for the description of navigation segments in interactive multi-view applications. Our work validates the potential of reducing the redundancies between multiple source a priori by jointly representing them, instead of relying on complex coding tools a posteriori. This perfectly fits with the concept of navigation segment recently introduced for interactive schemes.

The paper is organized as follows. Section 2 describes the concept of extended layered depth image and the steps needed to achieve the same. Section 3 describes the compression method for the extended layered depth image and Section 4 describes the experimental results.

\section{Extended layered depth image}

We propose here to describe all the information corresponding to the navigation segment by a single data structure, which represents all the points of the navigation segment while avoiding inter-view redundancies, i.e., 3D points are represented once and only once in contrary to the common multiview coding format. The proposed solution is an extension of the layered depth image format (LDI) \cite{Maugey_T_2013_tip_nav_dpimi}. LDI consists in representing the occluded areas of an image as an additional layer, described by an image having the shape of the foreground object. The number of layers corresponds to the number of occlusion levels in the scene. For instance the first layer consists of all information that can be seen from the reference camera viewpoint. Hence, it is a dense image with foreground and background information. Then the next layer consists of any objects/scene information, which was hidden behind the objects  in the first layer. Hence it is much sparser than the first layer. The third layer is even sparser, as it captures an even further level of occlusion, and so forth. Each layer captures not only the color information but also the depth information.
In the common LDI format, the data is restricted to an area corresponding to the  size of the reference viewpoint image and all the information out of this area is discarded; the reconstruction for a certain view from LDI is done with help of a compensation module \cite{Yoon_S_2007_jvlsi_fra_rpmvvucldi}. 

\begin{figure}[htb]
 \centering
\centerline{\epsfig{figure= 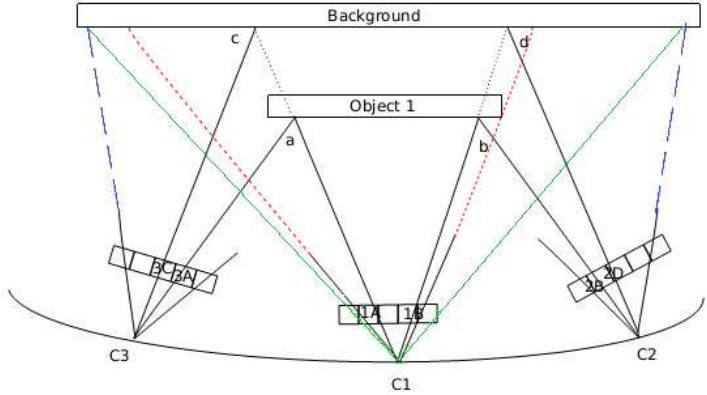,width=0.6\linewidth}}
\caption{Information represented by the extended LDI format (green solid line) and by the common LDI format (red dotted line).}
\label{fig:LDI}
\end{figure}

In our approach, all the information is captured in the original data representation by extending the size of the LDI. Hence, whereas the field of view is restricted to the reference camera in the original LDI, the field of view of the extended LDI is the maximum field of view captured by the original cameras in the navigation segment. We illustrate this important difference in Figure~\ref{fig:LDI}. The field of view is defined by the red dotted line, whereas for the new extended LDI, the field of view is defined by the green solid line, which is extended compared to the red dotted line with the aid of the fields of view of cameras C2 and C3 (blue dashed line).

We propose to extend the size of the LDI by projecting the boundary points of all the views to the LDI view positioned on the reference camera coordinates. We calculate the minimum and maximum distance of these warped points to the field of view of the reference camera and we adjust the size of the new LDI accordingly. The initial warping is done with respect to the coordinate system of the original LDI.  The size of the extended LDI is changed and we  incorporate also the pixels which are initially outside the boundaries of original LDI. Hence the (0,0) of the new coordinate system has gone to the leftmost point available and the warped points are represented with respect to the new coordinate system. The camera centre is finally adjusted according to the new coordinate system corresponding to the extended LDI. 

The extended LDI represents all information in a navigation segment and enables independent user navigation in the corresponding set of viewpoints. In other words this representation fully represents a particular navigation segment without any need of side information. Once a user moves to an area corresponding to another navigation segment, the extended LDI representation corresponding to the new navigation segment is transmitted to the user. The user can then freely navigate in the new segment, where any view can be generated by unwarping the corresponding extended LDI data.

The main advantage of the extended LDI format resides in the absence of redundancies at the representation level. A single representation captures all the available information corresponding to a navigation segment. Furthermore, it has a simple structure (close to a typical image format), which permits to use efficient existing image processing tools as opposed to more complex approaches such as mesh-based techniques that are necessary in geometry-based representations. In the next section we propose a compression algorithm for this new representation.

\section{Compression of extended LDI}

As seen in the last section, the extended LDI format helps to get rid of the geometrical redundancies (i.e., the repetition of the same 3D point in multiple views). We however still have to handle the inherent intra-view correlation for reducing the size of the data. It is important to remark that the extended LDI is not far from an image-based representation with additional depth-based information and multiple layers. Hence, classical image compression based methods with specific  modifications  for  the LDI structure can be used, as shown in \cite{Duan_J_2003_tip_com_ldi}. 
LDI has a special data structure, so that the classical compression tools cannot be directly applied. In particular, one needs to take care of three particular properties: i) there are multiple layers in LDI, ii) back layers are sparse and, iii) each pixel consists of multiple attributes. 

In order to adapt classical image compression methods, first of all every layer is handled separately. The literature in color image compression shows that multiple color components, such as the Y, Cr or Cb components can be separately compressed in an efficient manner as they are not highly correlated.  It is shown in paper \cite{Duan_J_2003_tip_com_ldi} that the same holds  for the LDI data. 
Hence each layer is divided into component images corresponding to Y, Cr, Cb and  depth values available at various pixel positions of a particular layer. It is observed that the component images of LDI at different layers are sparse and do not have a rectangular support. The sparsity increases as we go further back into the layers. Direct transform coding on such kind of data set is not very efficient.

Some kind of preprocessing needs to be done before applying the classical compression tools amongst the component images of different layers. One of the possible methods of preprocessing is to use data aggregation where all the pixels of the same row are pushed to the left of the row. The aggregated data is then transformed and encoded. Although transform efficiency is improved with aggregation,  vertical alignment is affected. Another problem with data aggregation is that the color distribution of the resulting aggregated images corresponding to different layers are very different from each other \cite{Yoon_S_2006_ppcs_cod_ldirmvv}. Hence the coding efficiency is not good because of the difficulty in prediction among the aggregated images.

As suggested in \cite{Yoon_S_2006_ppcs_cod_ldirmvv} in order to solve the limitations of simple data aggregation, the empty pixel locations of all the layer images can rather be filled using corresponding pixels in the first layer. This increases the prediction accuracy at the step of compression. So we use layer filling as the preprocessing method instead of aggregation.

After layer filling, we compress the resulting data. We use an alternative approach where we first pad the component image to a rectangular image, and then compress it with a rectangular still image coder such as JPEG-2000 \cite{jpeg2000}.  In the compression process, we have allocated more precision for the depth component compression, as errors in depth can lead to more serious errors. The bpp for depth has been chosen to be twice as that of color signal.

As there is a preprocessing step of layer filling involved,  a method is needed to revert the process of layer filling at decompression. It should keep only the desired information in the layers and remove the extra information added through the preprocessing step. This is achieved with the help of the number of layers (NOL) attribute at every spatial location in the LDI. NOL attribute describes  the number of layers that are present at a particular pixel location. With help of the NOL attribute, one can build a mask for every layer that determines if the information corresponding to a particular pixel position is present in the original layer or not. The mask $m_l(i,j)$ for layer l can be derived as \cite{Duan_J_2003_tip_com_ldi}:
\begin{equation}
m_l(i,j) = \left\{ \begin{aligned}
        1 \quad & n(i,j) > l \\
        0 \quad & n(i,j) \leq l
      \end{aligned} 
      \right.
\end{equation}
There is an LDI pixel at position (i,j) at layer l, if and only if $m_l(i,j)=1$.
 NOL image is compressed using JPEG-LS compression \cite{Weinberger_M_2000_tip_the_llicapsij} as it carries crucial information about the LDI structure and it needs to be preserved.

\section{Experiments}

In this section, we  study the performance of our solution from the perspective of interactive navigation applications using navigation segment descriptions. More precisely, we consider a navigation segment made of $N$ viewpoints. The optimized size of a navigation segment has been studied in \cite{Maugey_T_2013_tip_nav_dpimi}. Our purpose is to represent and code the information needed to recover these $N$ views with a maximum quality and minimum coding rate. The rate corresponds to the size (in kbs) of the information describing the navigation segment which permits the user to navigate in these $N$ views.

 \begin{figure}[htb]

\begin{minipage}{0.24\linewidth}
\centerline{\epsfig{figure= 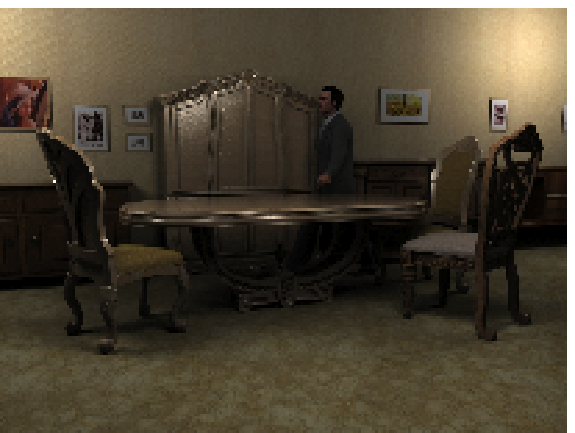,width=0.85\linewidth}}
\centerline{(a) camera 1}
\end{minipage}
\hfill
\begin{minipage}{0.24\linewidth}
\centerline{\epsfig{figure= 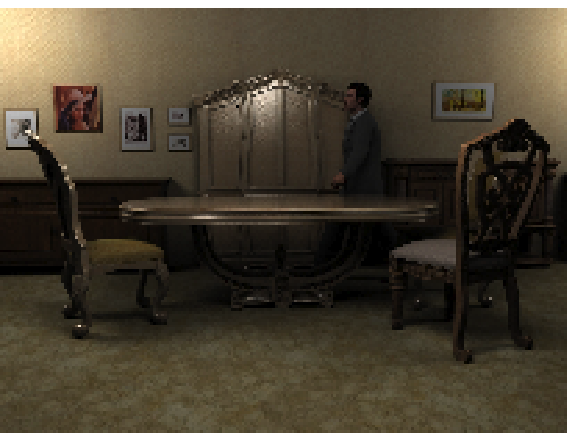,width=0.85\linewidth}}
\centerline{(b) camera 2}
\end{minipage}
\hfill
\begin{minipage}{0.24\linewidth}
\centerline{\epsfig{figure= 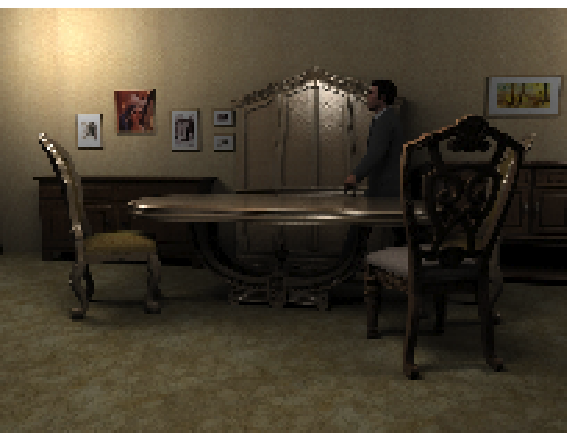,width=0.85\linewidth}}
\centerline{(c) camera 3}
\end{minipage}
\hfill
\begin{minipage}{0.24\linewidth}
\centerline{\epsfig{figure= 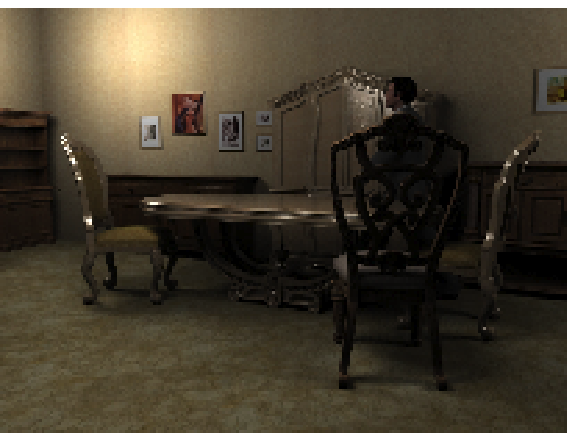,width=0.85\linewidth}}
\centerline{(d) camera 4}
\end{minipage}
\caption{Original Multiview images of ``Table" dataset, made of four viewpoints.}
\label{fig:ex_dataset}
\end{figure}

 \begin{figure}[htb]
\begin{minipage}{0.35\linewidth}
\centerline{\epsfig{figure= 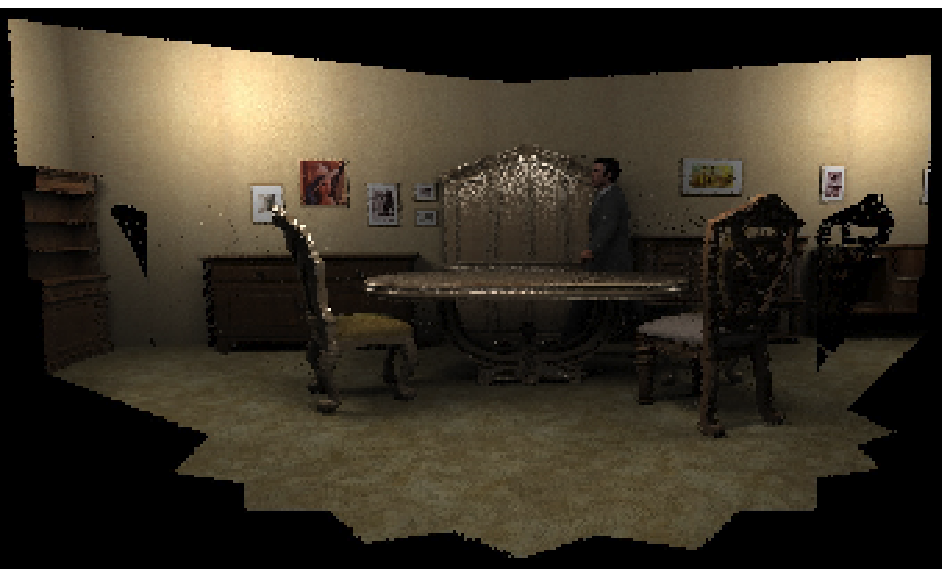,width=0.9\linewidth}}
\centerline{(a) layer 1}
\end{minipage}
\hfill
\begin{minipage}{0.31\linewidth}
\centerline{\epsfig{figure= 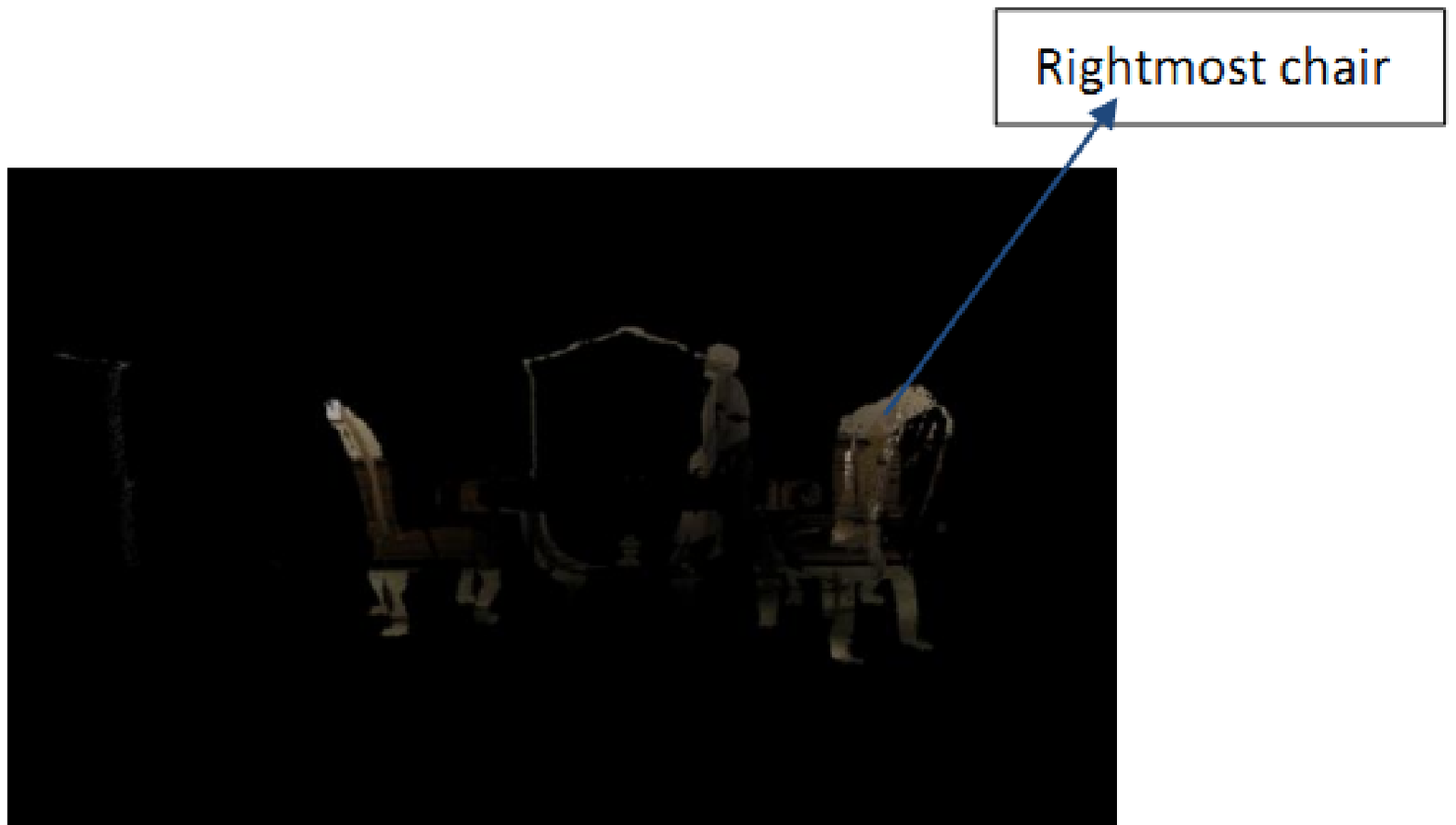,width=1\linewidth}}
\centerline{(b) layer 2}
\end{minipage}
\hfill
\begin{minipage}{0.27\linewidth}
\centerline{\epsfig{figure= 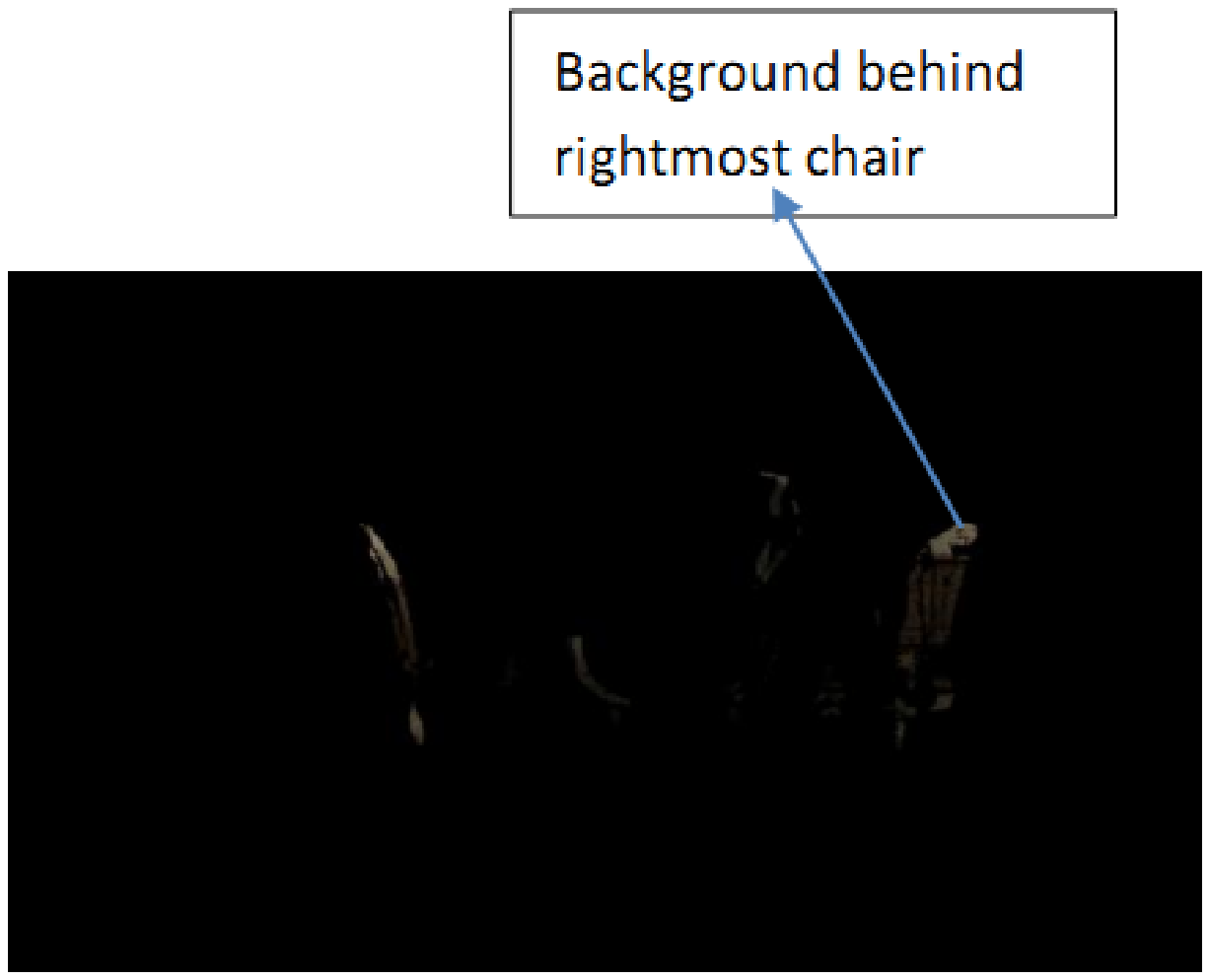,width=0.9\linewidth}}
\centerline{(c) layer 3}
\end{minipage}
\caption{Three layers of extended layered depth image format with reference on camera 2. }
\label{fig:ex_LDI}
\end{figure}

In this section, we want to validate the following idea: gathering multiple viewpoints in a reference layered depth image is a valid representation method in terms of flexibility. First of all, it is clear that, in the absence of compression, the LDI format permits to  reconstruct views with a high PSNR. We run our extended LDI method and compare it with the classical multiview representation \cite{jmvm}. For calculating the rate of the extended LDI, the compressed color information, compressed depth maps of different layers and coded mask are taken into account. The multiview representation strategy classically consists in describing the scene with multiple images (with their depth) and in coding them with a classical multiview video coding (MVC) tool. More complex tools exist nowadays. However, even if they reach better performance, they still rely on the idea of removing the multiple occurrence of 3D pixels by predictive coding, which is not suited for user's interactivity. Hence we introduce a new form of data representation, and we limit our rate-distortion performance comparisons to the well-known MVC standard as our purpose is to show that the flexibility introduced in our novel data representation does not necessarily comes at the price of compression penalty. The number of images $M$ involved in the representation  is generally fixed by the number of captured images. If the navigation segment contains virtual viewpoints, it means that $M<N$. In other words, the $M-N$ virtual frames are recovered with depth-based interpolation tools (depth image-based rendering techniques \cite{Tian_D_2009_pspie_vie_sttdv}) using the $M$ decoded viewpoints. This corresponds to the classical approach adopted in most of the standards. For calculating the rate of this multiview image representation, we calculate the size of both the compressed color map and the compressed depth map of the $M$ original views. The fundamental difference between the proposed extended LDI approach and the traditional multiview image representation (designed by MVC in the following) is, as mentioned before, the fact that the geometrical redundancy is either removed at the representation level or at the coding step. It means that our extended LDI technique does not code $M$ viewpoints, but rather gathers the $M$ viewpoints in one signal data (both schemes then interpolate the $N-M$ other ones at decoder side). The following experiments compare these two conceptual approaches.

\begin{figure}[htb]
 \centering
 \begin{minipage}{0.32\linewidth}
\centerline{\epsfig{figure= 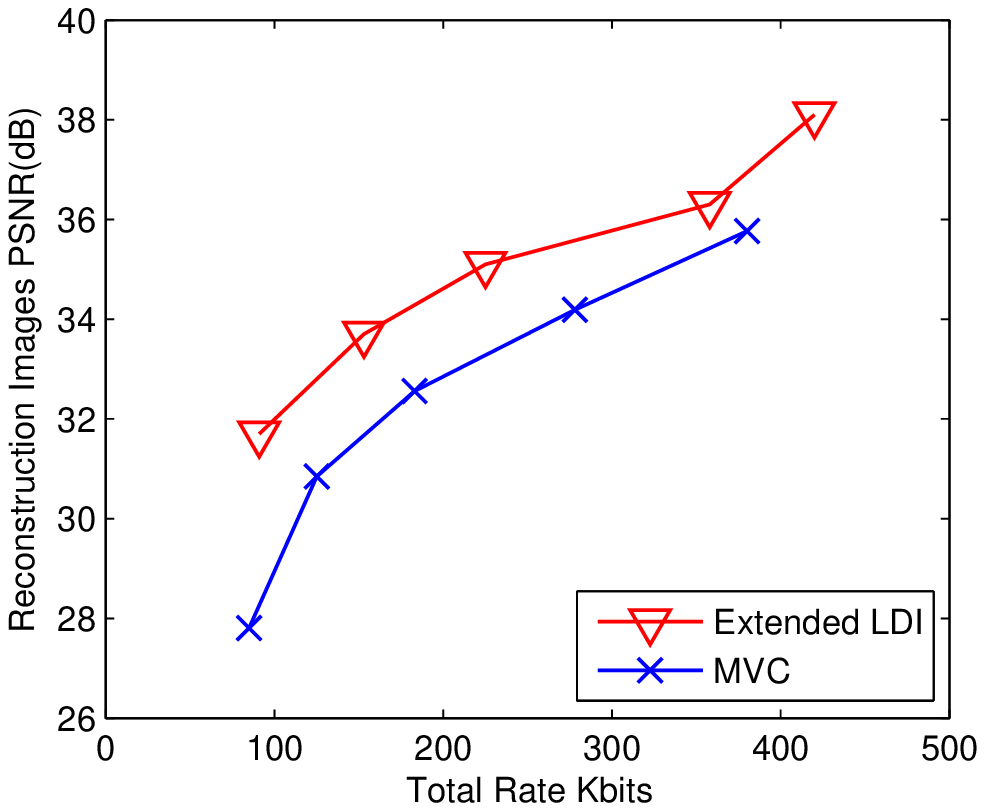,width=1\linewidth}}
\centerline{(a)}
\end{minipage}
\hfill
 \begin{minipage}{0.32\linewidth}
\centerline{\epsfig{figure= 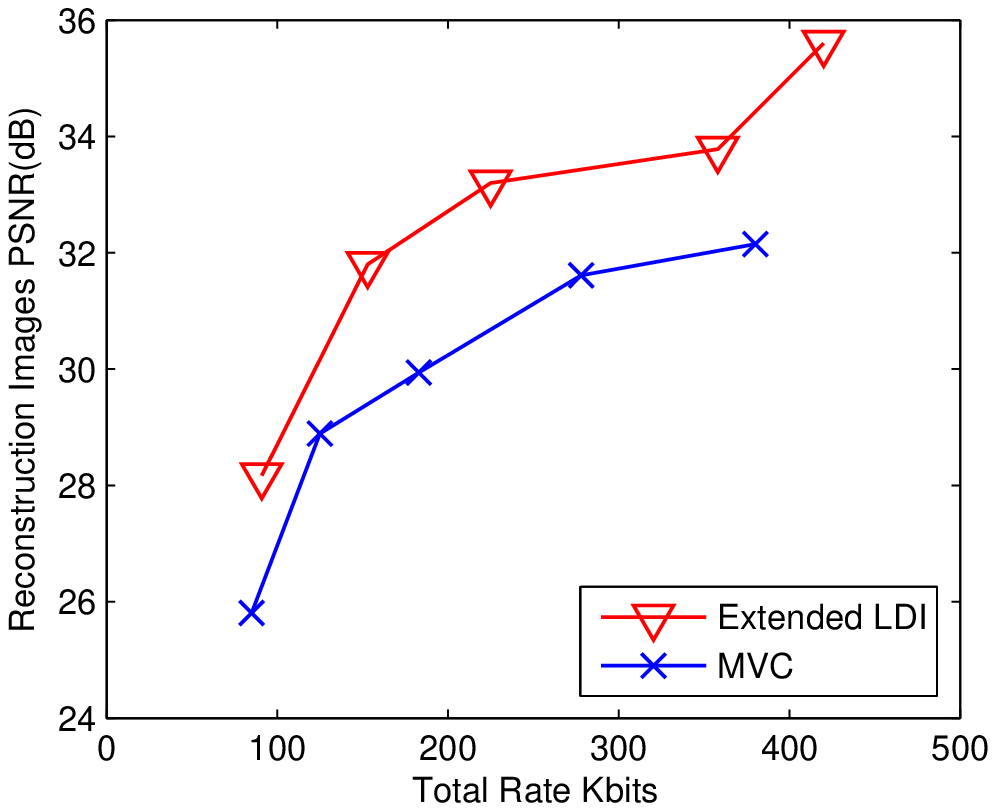,width=1\linewidth}}
\centerline{(b)}
\end{minipage}
 \begin{minipage}{0.32\linewidth}
\centerline{\epsfig{figure= 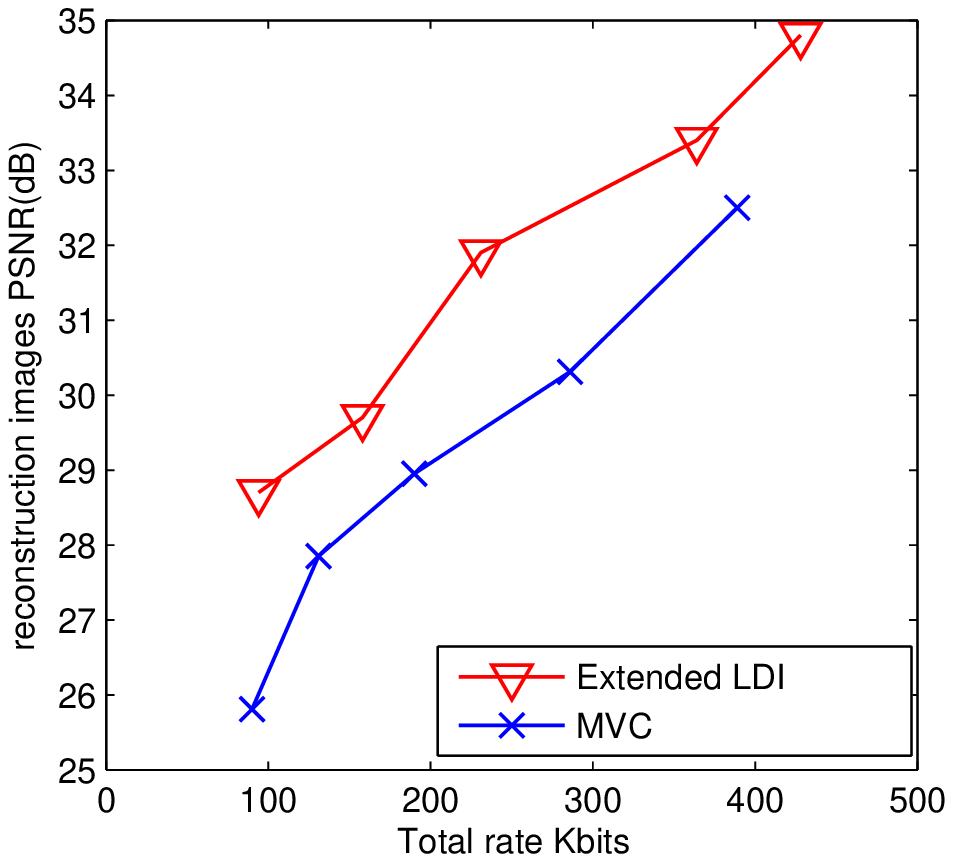,width=0.9\linewidth}}
\centerline{(c)}
\end{minipage}
\caption{Rate distortion performance of extended LDI and multiview images representations for (a) ``Table", (b) ``Breakdancers" and (c) ``Ballet"  datasets. PSNR values are averaged over $15$ viewpoints of the navigation segment.}
\label{fig:exp}
\end{figure}

For designing the navigation segments, we use the multiview images ``Table" depicted in Figure~\ref{fig:ex_dataset}  and the ``Breakdancers" sequence \cite{Zitnick_C_2004_acmtg_hig_qvviulr}. In order to measure the performance in terms of distortion, we first synthesize $N-M$ virtual views from the original $M$ images to generate ground truth data. These are generated by first interpolating the camera parameters and then using depth-based warping from the two neighboring views  \cite{Tian_D_2009_pspie_vie_sttdv}.  We then measure the distortion of the reconstructed $N-M$ virtual views at decoder with respect to this ground truth data. For both ``Table" and ``Breakdancers" image sets, $M$ is equal to $4$ and $N$ is equal to $15$. Moreover, the three layers of extended LDI are  obtained after warping information from all the cameras to the viewpoint of camera $2$ for ``Table" and $3$ for ``Breakdancer" and ``Ballet". An example of extended LDI is depicted in Figure~\ref{fig:ex_LDI} for the ``Table" dataset. 
Results for both datasets are depicted in Figure~\ref{fig:exp}. We observe that, the extended LDI representation has a better performance  than the traditional MVC format. It validates the idea of removing the redundancies at the representation level. This is very interesting for navigation applications since it reduces the size of a navigation segment sent to users when they aims at navigating between the $N$ views of this segment. It also validates the idea that multiview images should not be simply considered as an extension of temporal 2D video, and should rather be taken into account as a real 3D signal with geometrical information.

\begin{figure}[htb]
 \centering
 \begin{minipage}{0.48\linewidth}
\centerline{\epsfig{figure= 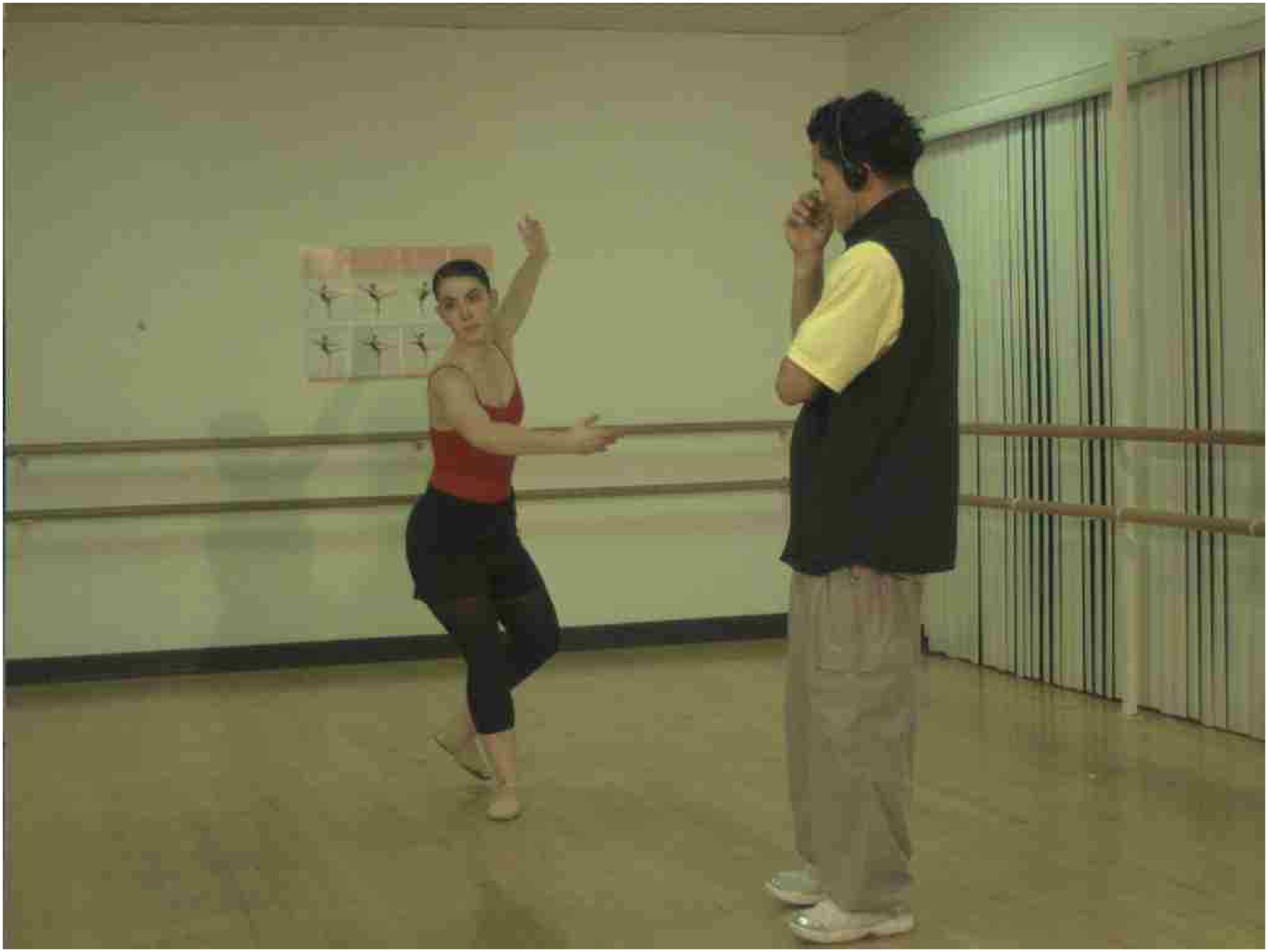,width=1\linewidth}}
\centerline{(a)}
\end{minipage}
\hfill
 \begin{minipage}{0.48\linewidth}
\centerline{\epsfig{figure= 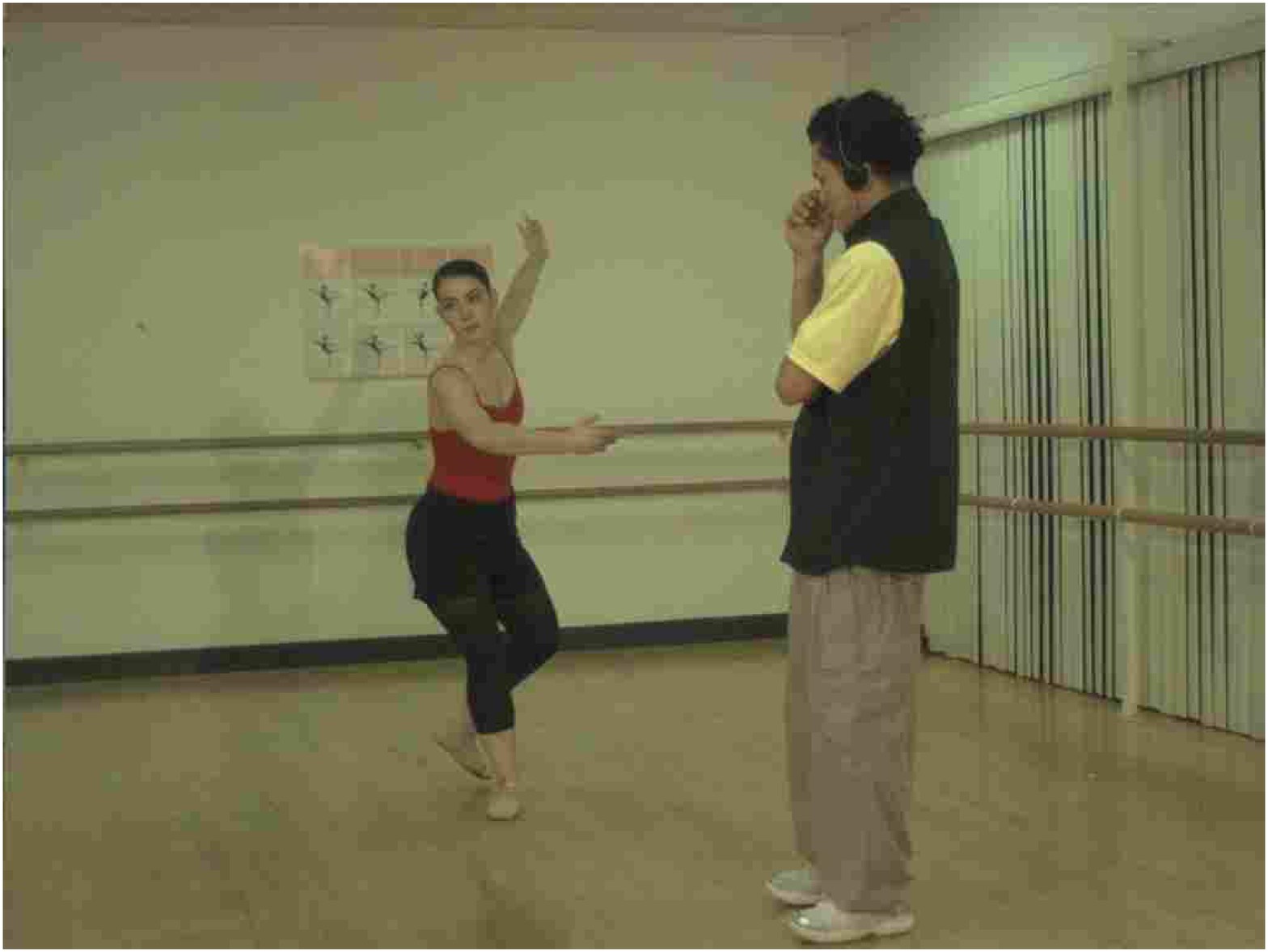,width=1\linewidth}}
\centerline{(b)}
\end{minipage}
\hfill
 \begin{minipage}{0.6\linewidth}
\centerline{\epsfig{figure= 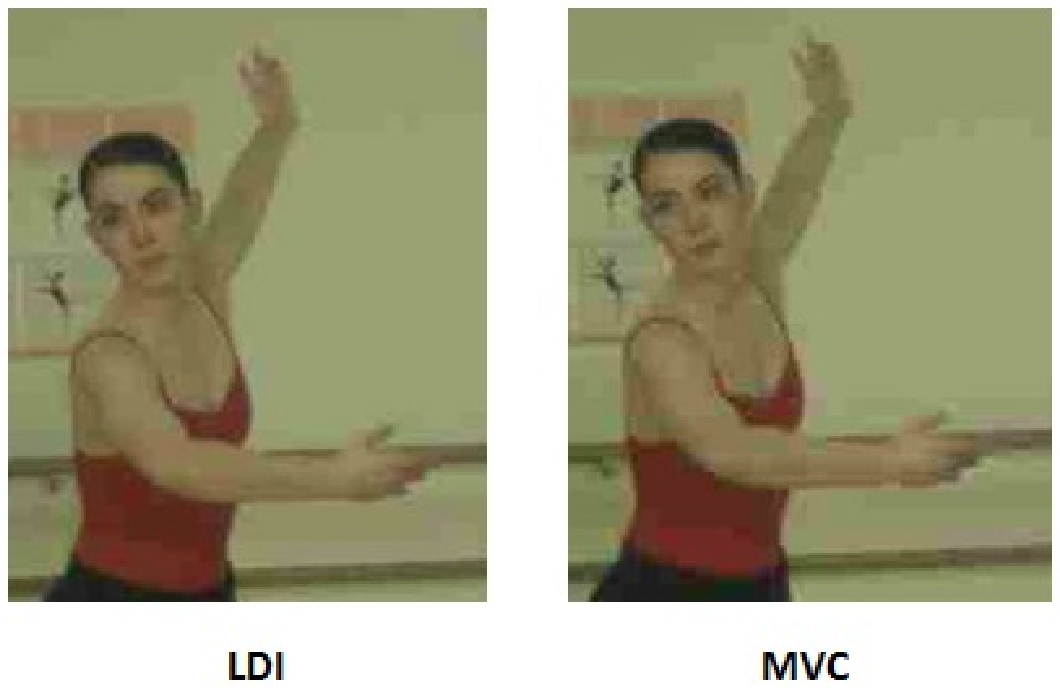,width=0.7\linewidth}}
\centerline{(c)}
\end{minipage}
\caption{Visual comparison for synthesized view at camera location 2.01 for ``Ballet" Sequence" between MVC and LDI formats.}
\label{fig:visualballet}
\end{figure}

Fig.~\ref{fig:visualballet} depicts the visual results obtained from an experiment on ``Ballet" sequence by reconstructing a virtual view at camera location 2.01 from the compressed extended LDI and MVC. For the same total bit rate, we can see better visual results for the extended LDI format compared to MVC. Moreover, block-based schemes are not efficient when rotations occur between camera transitions. We see that for a similar bitrate, LDI solution works with a lower QP than the MVC, because of the compactness of the proposed representation.

\begin{figure}[htb]
 \centering
 \begin{minipage}{0.48\linewidth}
\centerline{\epsfig{figure= 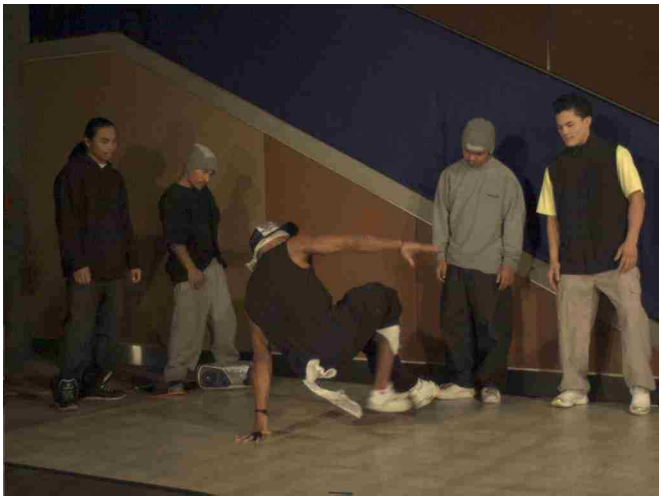,width=1\linewidth}}
\centerline{(a)}
\end{minipage}
\hfill
 \begin{minipage}{0.48\linewidth}
\centerline{\epsfig{figure= 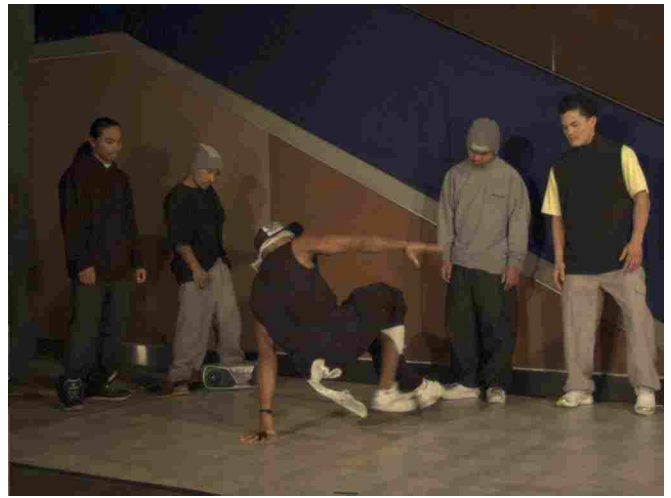,width=1\linewidth}}
\centerline{(b)}
\end{minipage}
\caption{Visual results for synthesized view at camera location 2.01 for the ``Breakdancers" sequence for (a) extended LDI (b) MVC}
\label{fig:visualbreakdancers}
\end{figure}

\begin{figure}[htb]
 \centering
 \begin{minipage}{0.48\linewidth}
\centerline{\epsfig{figure= 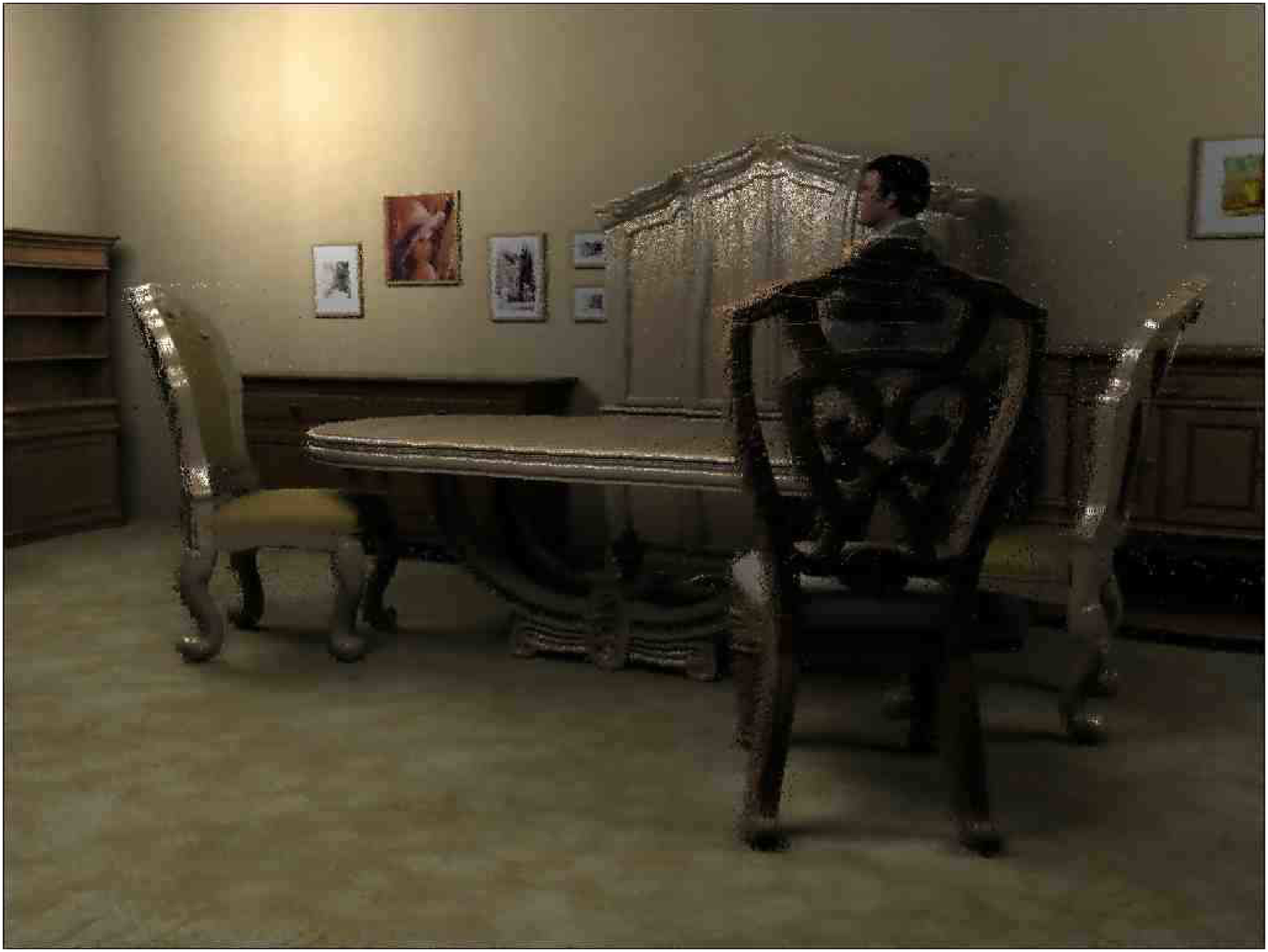,width=1\linewidth}}
\centerline{(a)}
\end{minipage}
\hfill
 \begin{minipage}{0.48\linewidth}
\centerline{\epsfig{figure= 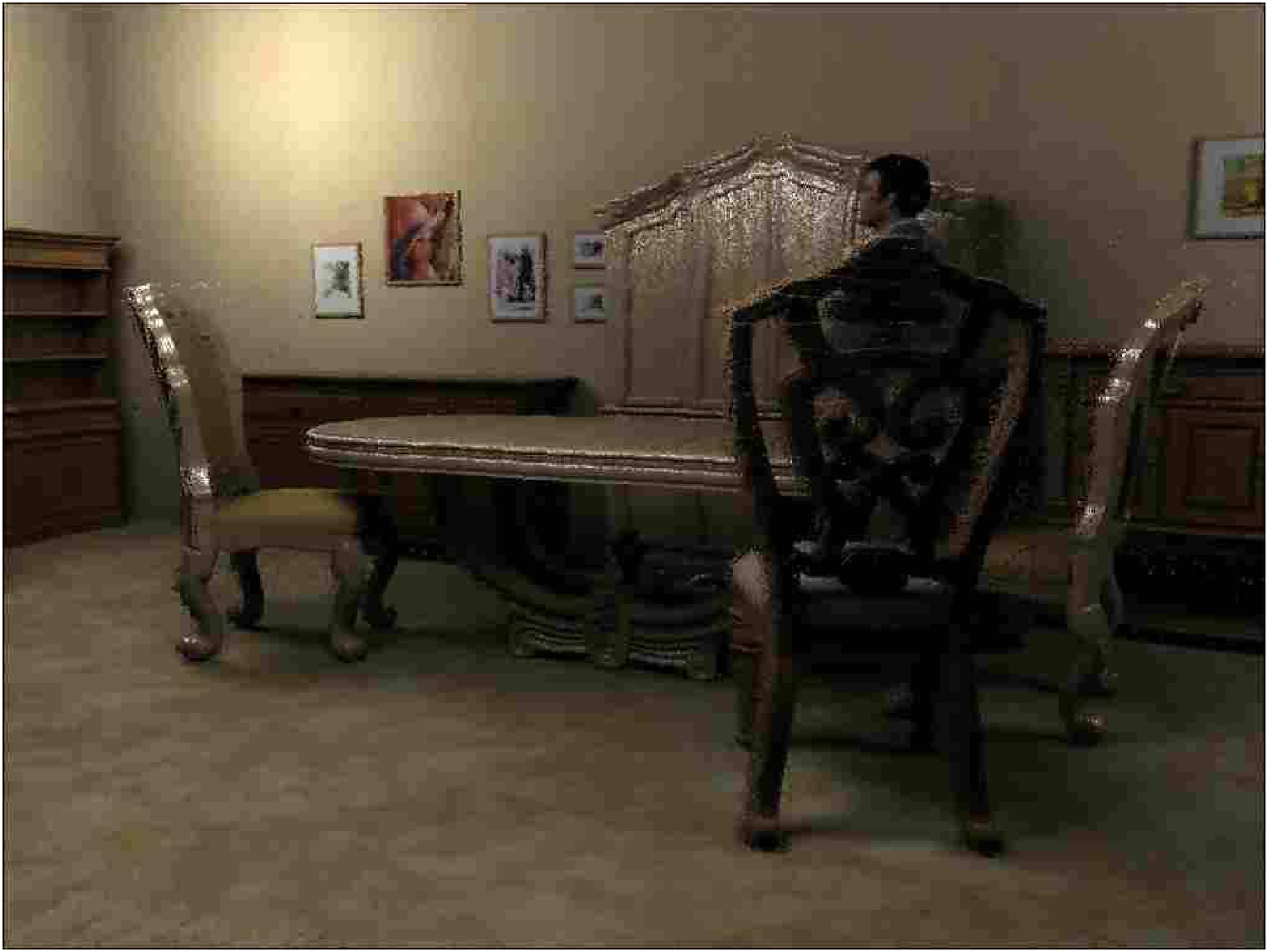,width=1\linewidth}}
\centerline{(b)}
\end{minipage}
\caption{Visual results for synthesized view at camera location 1.01 for the ``Table" sequence for (a) extended LDI (b) MVC}
\label{fig:visualtable}
\end{figure}

\section{Conclusion}
We have proposed a new representation and coding method that is appropriate for describing navigation segments in interactive multiview applications. It is based on a layered depth image representation, which has the advantage to avoid redundancy in the multiview information representation. We propose an extended layered depth image format, which describes every point in the navigation segment once and only once. We further provide an adapted compression strategy and we show through experiments that our new data format provides interesting rate distortion performance in the context of emerging interactive applications. 

\IEEEpeerreviewmaketitle

\ifCLASSOPTIONcaptionsoff
  \newpage
\fi

\bibliographystyle{IEEEtran}
\bibliography{/Users/thomasmaugey/Documents/EPFL/bibli/abbr,/Users/thomasmaugey/Documents/EPFL/bibli/epfl}

\end{document}